\documentclass[11pt,letterpaper]{article}

\usepackage[utf8]{inputenc}
\usepackage[left=1in,right=1in,top=1in,bottom=1in]{geometry}
\usepackage{amsmath}
\usepackage{amsfonts,amssymb} 
\usepackage{lmodern}
\usepackage{siunitx}
\usepackage{physics}
\usepackage{comment}
\usepackage{placeins}
\usepackage{cite}
\usepackage[affil-it]{authblk}
\usepackage{rotating}
\usepackage{pdflscape}
\usepackage{fancyhdr}
\usepackage{multirow}
\usepackage{hyperref}
\usepackage{caption}
\usepackage{subcaption}
\usepackage{epsfig,epstopdf}
\usepackage{graphicx,color}
\usepackage[colorinlistoftodos,prependcaption,textsize=tiny]{todonotes}
\usepackage{lineno}
\usepackage{enumitem}
\usepackage{xcolor}
\usepackage{xspace}

\hypersetup{
    colorlinks=true,
    linkcolor=blue,
    filecolor=blue,
    urlcolor=blue,
    citecolor=blue
}

\newcommand{\pt}{\ensuremath{p_{T}}\xspace}

\newcommand\snowmass{
    \begin{center}\rule[-0.2in]{\hsize}{0.01in} \\
        \rule{\hsize}{0.01in} \\
        \vskip 0.1in Submitted to the Proceedings of the US Community Study \\ 
        on the Future of Particle Physics (Snowmass 2021) \\ 
        \rule{\hsize}{0.01in} \\
        \rule[+0.2in]{\hsize}{0.01in} 
    \end{center}
}

\title{\bf 4-Dimensional Trackers}
\date{\today}

\author[1]{Doug Berry} %rberry@gmail.com
\author[2]{Valentina Cairo}
\author[3]{Angelo Dragone}
\author[4]{Matteo Centis-Vignali} % mcentisvignali@fbk.eu
\author[5]{Gabriele Giacomini} % giacomini@bnl.gov
\author[1]{Ryan Heller}
\author[1]{Sergo Jindariani} %sergo@fnal.gov
\author[6]{Adriano Lai} %adriano.lai@ca.infn.it
\author[2]{Lucie Linssen} % Lucie.Linssen@cern.ch
\author[1]{Ron Lipton} %lipton@fnal.gov
\author[1]{Chris Madrid}
\author[3]{Bojan Markovic} %markovic@slac.stanford.edu
\author[7]{Simone Mazza}
\author[7]{Jennifer Ott} %jeott@ucsc.edu
\author[3]{Ariel Schwartzman}
\author[8]{Hannsj\"org Weber} % hannsjoerg.weber@physik.hu-berlin.de
\author[9]{Zhenyu Ye} % yezhenyu@uic.edu

\affil[1]{Fermi National Accelerator Laboratory, Batavia, IL 60510, USA}
\affil[2]{CERN, Conseil Europ\'{e}en pour la Recherche Nucl\'{e}aire, 1211 Geneva 23, Switzerland}
\affil[3]{SLAC National Accelerator Laboratory; Menlo Park, California 94025, USA}
\affil[4]{Fondazione Bruno Kessler, Trento, Italy}
\affil[5]{Brookhaven National Laboratory, Upton, 11973, NY, USA}
\affil[6]{Istituto Nazionale Fisica Nucleare, Sezione di Cagliari, Cagliari, Italy}
\affil[7]{SCIPP, University of California Santa Cruz, Santa Cruz, CA 95064, USA}
\affil[8]{Institut f\"ur Physik, Humboldt-Universit\"at zu Berlin, 12489 Berlin, Germany}
\affil[9]{University of Illinois at Chicago, Chicago, IL 60607, USA}

\begin{document}

\snowmass

{\let\newpage\relax\maketitle}

\maketitle

\noindent\textbf{Contact Information:}\\
Valentina Cairo (valentina.maria.cairo@cern.ch) \\
Ryan Heller (rheller@fnal.gov)\\
Simone Mazza (simazza@ucsc.edu)\\
Ariel Schwartzman (sch@slac.stanford.edu)\\

%\vspace{2cm}
%\hline
%\vspace{0.4cm}
%{\bf Abstract:} ... 
%\vspace{0.4cm}
%\hline

%\thispagestyle{empty}
%\newpage
\setcounter{page}{1}

\section{Introduction}
%Physics motivations for 4D trackers
%Resolutions (space-time), layout considerations, constrains: Power and material budget)

Precision timing at the level of 10-30ps will be a game-changing capability for detectors at future collider experiments. For example, the ability to assign a time stamp with 30ps precision to particle tracks will allow to mitigate the impact of pileup at the High-Luminosity LHC (HL-LHC). With a time spread of the beam spot of approximately 180ps, a track time resolution of 30ps allows for a factor of 6 reduction in pileup. 

Both ATLAS and CMS will incorporate dedicated precision timing detector layers for the HL-LHC upgrade~\cite{CERN-LHCC-2020-007, CMS:2667167}. Timing information will be even more important at future high-energy, high-luminosity hadron colliders with much higher levels of pileup. At a future 100 TeV p-p hadron collider, for example, one of the primary challenges will be the efficient reconstruction of charged particle tracks in an environment of unprecedented pileup density. A powerful way to address this challenge is by fully integrating timing with the 3-dimensional spatial information of pixel detectors. An integrated 4-dimensional tracker with hit timing resolution at the levels of $\sim$10ps can drastically reduce the combinatorial challenge of track reconstruction at extremely high pileup densities~\cite{Sadrozinski_2017}.  

While timing information will be critical to mitigate the impact of pileup, it is not the only way in which it will enhance the event reconstruction of future hadron and lepton colliders. Timing information offers completely new handles to detect and trigger on long-lived particles (LLP)~\cite{CMS:2667167, Liu_2019}, expand the reach to search for new phenomena~\cite{CMS:2667167}, and enable particle-ID capabilities for pion/kaon separation at low transverse momentum~\cite{CMS:2667167}. 4D devices with coarse timing capabilities at $\sim$ns level but with similar granularity as regular tracking devices at the other end of 4D phase space can complement the fast timing layers for an enhanced overall 4D tracking. 

The optimal design of future 4D trackers will involve three key considerations: sensors with adequate spatial and time resolution, low power and low noise readout electronics, and overall detector layout, including material considerations. Significant research and development effort is required to understand how to best design 4D trackers and how all these aspects will impact physics performance.   

The following sections describe specific considerations for the integration of timing within tracking detectors at various future collider detectors and upgrades of existing experiments.

\subsection{Hadron colliders}

The High Luminosity phase of the LHC (HL-LHC), expected to start around 2029, will be the first collider where precision timing capabilities will be necessary in order to accurately reconstruct events. The HL-LHC will feature approximately 200 nearly simultaneous collisions during each bunch crossing, dispersed over a few hundred of picoseconds. To disentangle  this complex environment, CMS and ATLAS are building precision timing layers that will timestamp each track with a precision of roughly 30 ps, enabling the separation of slightly out-of-time pileup tracks from the primary interaction. These timing detectors, the ATLAS High Granularity Timing Detector (HGTD) \cite{CERN-LHCC-2020-007} and the CMS MIP Timing Detector (MTD) \cite{CMS:2667167} in the forward regions will be based on novel Low Gain Avalanche Detectors (LGADs), thin silicon sensors with moderate internal gain. Though these sensors provide precise time information, they can only achieve modest spatial segmentation of approximately \SI{1}{\milli\m}. Thus, the timing layers serve as precursors to true 4D trackers and as excellent first platforms for developing precision timing in collider environments.

The ATLAS and CMS trackers will also be upgraded to cope with the extreme conditions of the HL-LHC \cite{CERN-LHCC-2017-021, CERN-LHCC-2017-005}. Due to the high radiation dose in close proximity of the interaction point, the two innermost pixel layers of the ATLAS inner tracker (ITk) will need to be replaced during the course of the HL-LHC. The longer timescale of this replacement offers an additional opportunity to develop precision timing in the pixel layers as well, naturally complementing the HGTD in the forward region. Even a single pixel barrel layer with timing capabilities could achieve a vast performance improvement and would also be a striking first demonstration of true 4D tracker. The pixel time resolution needed to substantially boost the ATLAS performance is still to be evaluated, along with the impact of the material introduced in the tracker. Detailed simulation studies are needed to assess the potential of such detector layout and ultimately probe its feasibility.

In general, the additional information provided by precision timing at the HL-LHC will have substantial impact on some of the most challenging and important physics analyses, including searches for Higgs pair production, measurement of missing energy, searches for long-lived particles, and particle identification (PID) in heavy ion collisions.

Beyond HL-LHC, one of the key challenges in the design of the Future hadronic Circular Collider (FCC-hh \cite{FCC-hh-CDR}) which would follow the Future electron-positron Circular Collider described below (FCC-ee \cite{FCC-ee-CDR}) arises from a further increased number of pile-up events O(1000), almost an order of magnitude more than at the HL-LHC. Track-finding and identification of vertices based on traditional 3D algorithms will be extremely difficult, making a very clear case for the usage of 4D technology in all tracking layers. Timing information at each layer allows to associate hits that are consistent in time, which will dramatically reduce the complexity in the extreme occupancy environment. 

\begin{figure}
    \centering
    \includegraphics[width=0.8\linewidth]{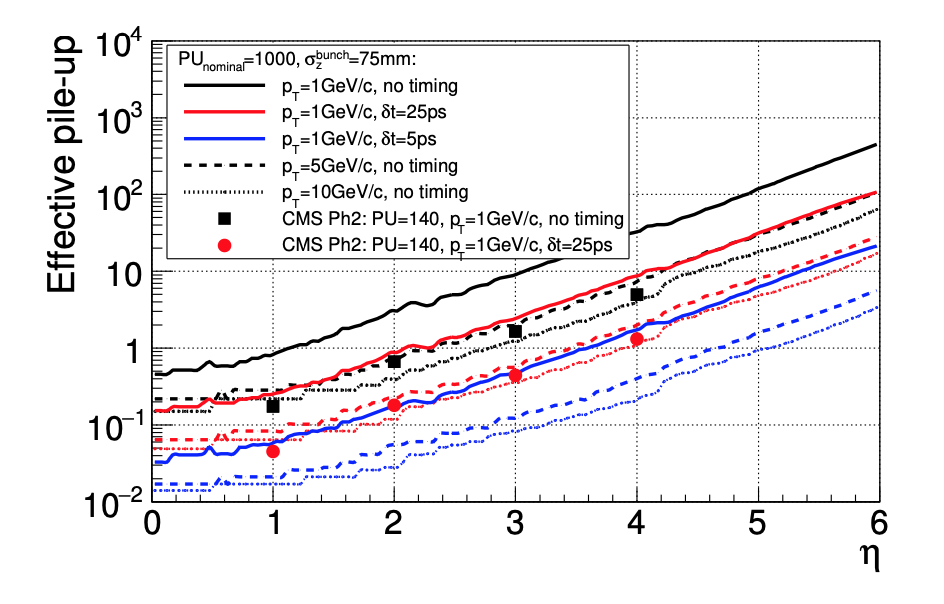}
    \caption{From Ref. \cite{Drasal:2674721}. An effective pile-up in the FCC-hh tracker. Several options of timing resolution per track in 3D vertexing are assumed: no timing (black), $\delta$ t = 25 ps (red) and $\delta$ t = 5 ps (blue). Several $p_{T}$ values are shown: 1 GeV/c (solid), 5 GeV/c (dashed) and 10 GeV/c (dotted). For reference the effective pile-up for CMS Phase 2 layout, $p_{T}$ = 1 GeV/c and nominal pile-up=140 is added.}
    \label{fig:FCC-pu}
\end{figure}

One important question to answer is what timing resolution is needed for a collider like the FCC. A key metric used is the effective pileup, which represents the mean number of vertices coincident with the primary interaction that cannot be resolved based on the available spatial or timing information. As shown in Fig. \ref{fig:FCC-pu}, extreme timing resolution of 5 - 10 ps per track is essential to keep the effective pileup low and prevent the merging of unrelated vertices.

%\footnote{Effective pile-up is defined as the number of pile-up vertices which effectively lead to a confusing assignment of low $p_{T}$ tracks to the original primary vertex} under control at large pseudorapidities ($|\eta|>3$) which would otherwise reach level of tens or hundreds leading to large merging effects in vertex reconstruction and large confusion in vertex selection.

At the same time, the higher energies and luminosities typical of the FCC-hh pose very stringent constraints on the detector design itself, for instance on the radiation hardness of the silicon modules, limiting either the lifetime of the inner detector or the suitable detector technologies.
In the layouts under study \cite{FCC-hh-CDR}, the beampipe is taken to have a radius of 20 mm. At the radius of the innermost tracker layers, 25 mm, the radiation levels expected after 
30 $ab^{-1}$ are of the order of a dose of 0.4 GGy and a fluence of $6 \times 10^{17}$ per $cm^{2}$ 1 MeV neq. 
These are approximately 30 times (600 times) more intense than the environment at the HL-LHC (LHC). No existing tracker technologies can satisfy these requirements, and dedicated R\&D efforts targeting extreme timing resolutions and radiation hardness must be pursued. The achieved time resolution and radiation hardness will, also in this case, be correlated with the spatial resolution and the changes in the material budget, thus analyzing the interplay among them is of key interest.

\subsection{Electron-Positron colliders}

A variety of future electron-positron $e^+e^-$ colliders is being investigated world-wide to identify a future Higgs Factory.
The proposals range from linear (International Linear Collider ILC \cite{behnke2013international}, Compact Linear Collider CLIC \cite{linssen2012physics}, Cool Copper Collider C$^3$ \cite{C3}) to circular (FCC-ee \cite{FCC-ee-CDR}, Circular Electron Positron Collider CEPC \cite{CEPC_1, CEPC_2}) $e^+e^-$ colliders.

In general, the usage of 4D tracking technology at $e^+e^-$ colliders is subject to very different conditions compared to that of experiments at hadron machines: both the numbers of collisions per beam crossing and the radiation levels are orders of magnitude lower, but the physics measurements are normally targeting very high precision, imposing track parameter resolutions to be extremely good, thus requiring very low passive material in the vertexing and tracking detectors.

Most of the studies performed so far focus on the usage of time at the ns resolution level as part of the object reconstruction chain, while studies of potential applications of precision timing at the ps level are still to be further investigated in $e^+e^-$ colliders. 

The ILC is a proposed 20 km $e^+e^-$ linear collider at the energy frontier, with an initial baseline center-of-mass energy of 250 GeV. Two detector concepts have been studied at the ILC: the Silicon Detector (SiD) \cite{SiDLOI,breidenbach2021updating} and the International Large Detector (ILD) \cite{ILDLOI,behnke2013international,ILD}.

SiD is a compact detector based on a powerful silicon pixel vertex detector, silicon tracking, silicon-tungsten electromagnetic calorimetry (ECAL) and highly segmented hadronic calorimetry (HCAL). SiD also incorporates a 5T solenoid, iron flux return, and a muon identification system.
The choice of silicon detectors for tracking and vertexing ensures that SiD is robust with respect to beam backgrounds or beam loss, provides superior charged-particle momentum resolution, and eliminates out-of-time tracks and backgrounds.

While timing layers with resolutions at the level of the nanosecond could be used in the HCAL to help suppress backgrounds, the recent developments in fast-timing detectors could bring in improvements to the SiD layout, as described in Ref. \cite{breidenbach2021updating}. 

%Fig. \ref{fig:HitTime_SiD} shows the timing distribution of the beam background hits displaying the collision and then with a clear separation the backgrounds hits from a back-splash from the forward instrumentation.

%\begin{figure}
%    \centering
%    \includegraphics[width=0.7\linewidth]{HitTime_SiD.pn%g}
%    \caption{From Ref. \cite{breidenbach2021updating}. % The time distribution of beam background hits in the SiD %Vertex Detector Endcap.}
%    \label{fig:HitTime_SiD}
%\end{figure}

\begin{figure}
    \centering
    \includegraphics[width=0.7\linewidth]{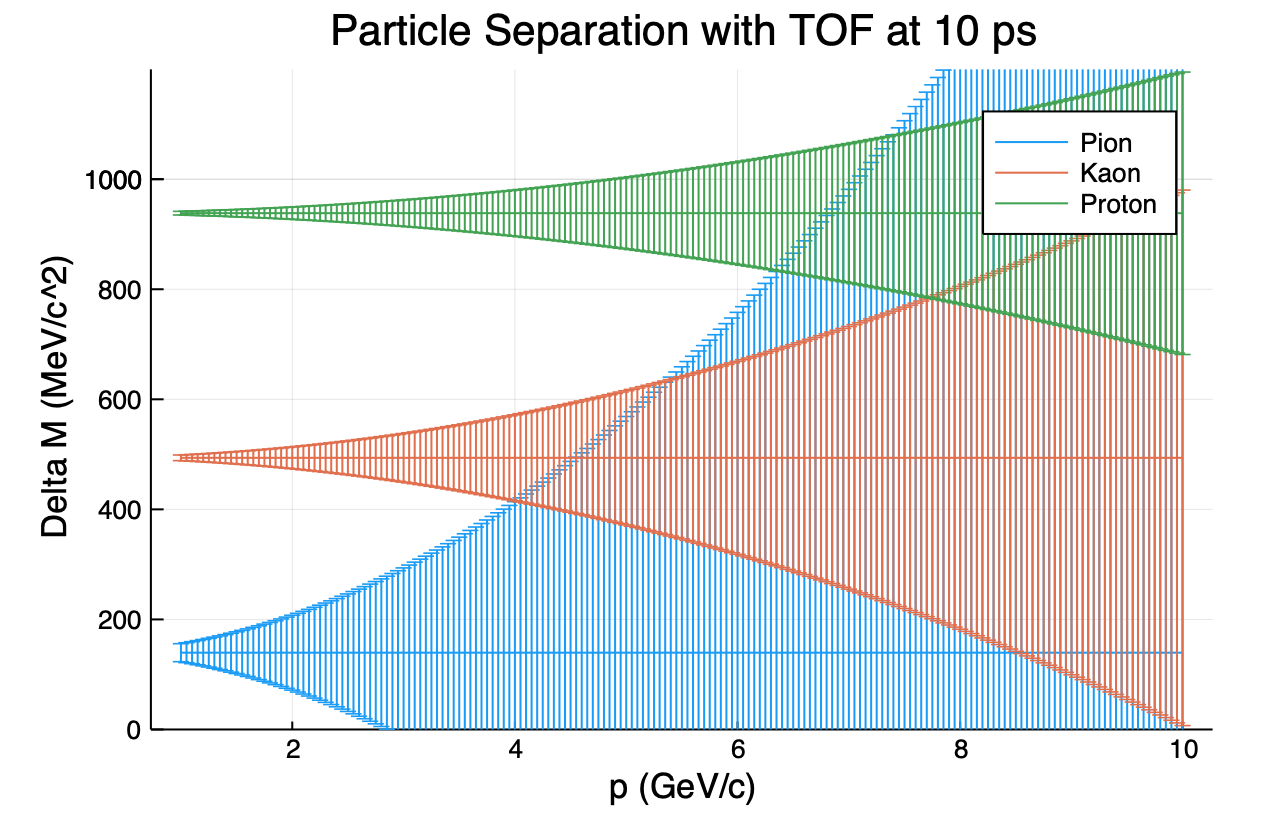}
    \caption{From Ref. \cite{breidenbach2021updating}. Mass resolution for a time-of-flight system with a performance of 10 ps in SiD.}
    \label{fig:TOF_SiD}
\end{figure}

Time resolutions at the ps level would, for example, make it possible to exploit time-of-flight (TOF) for low-momentum particle identification (PID) if timing layers were added to the tracking system or in between the tracker and the ECAL. Fig. \ref{fig:TOF_SiD} shows that, in SiD, a TOF system with time resolution of 10~ps allows for PID up to a momenta of a few GeV. 

The other dector concept at the ILC, i.e. ILD, has been designed as a multi-purpose detector for optimal particle-flow (PFA) performance. Its tracking systems differs from the SiD one: a high-precision vertex detector is followed by a hybrid tracking system, realised as a combination of silicon tracking with a time-projection chamber (TPC). The complete system, along with a calorimeter, is located inside a 3.5~T solenoid. 

\begin{figure}
    \centering
    \includegraphics[width=0.5\linewidth]{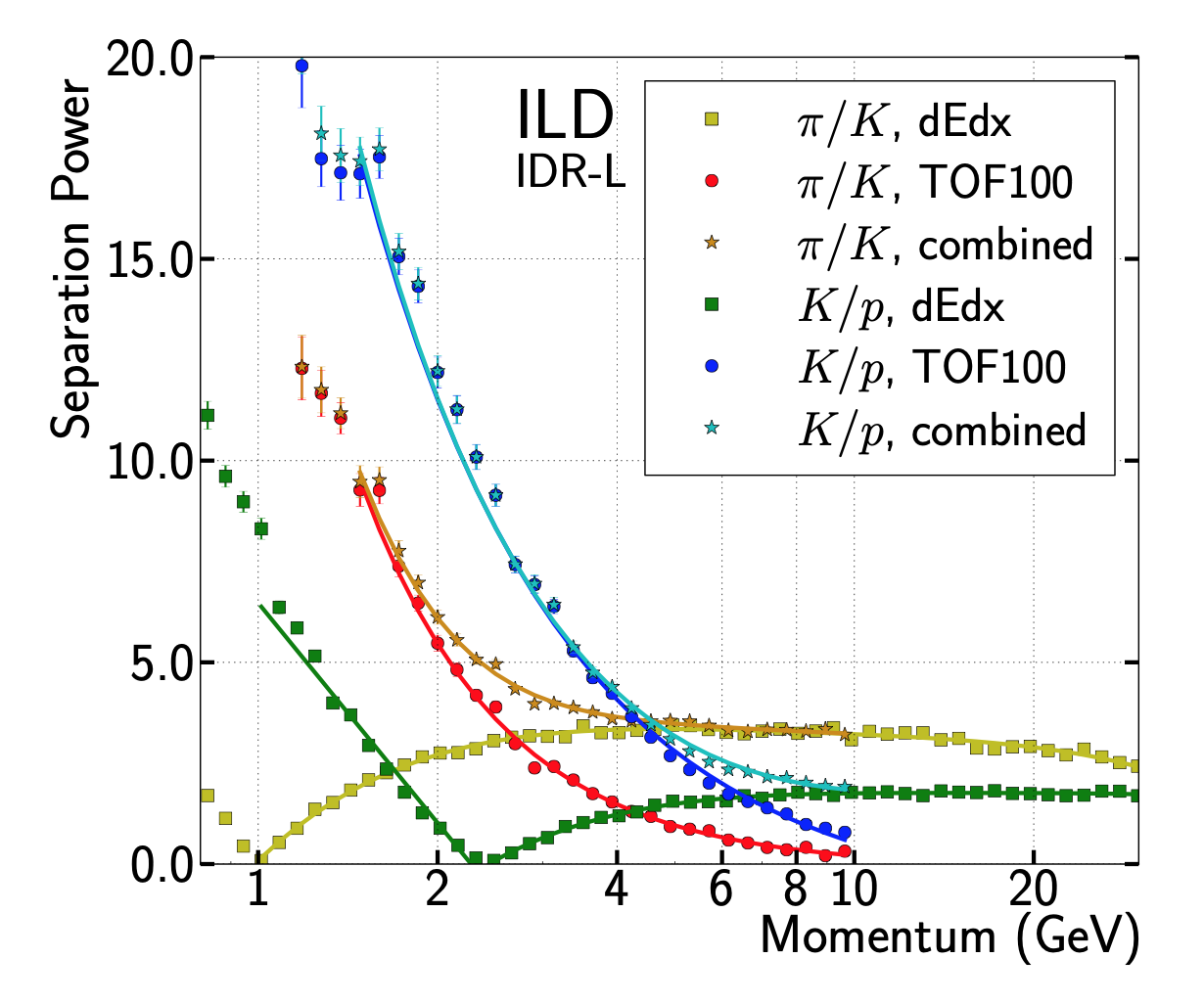}
    \caption{From Ref. \cite{theildcollaboration2020international}.  Particle separation power for $\pi/k$ and $K/p$ based on the dE/dx measurement in the TPC and on a time-of-flight estimator from the first ten ECAL layers. The separation power obtained when the information from the two systems is combined is also shown.}
    \label{fig:dEdx_TOF_ILD}
\end{figure}

Unlike  SiD, the current ILD design already has PID capabilities enabled by measurements of the energy loss of charged particles due to ionisation ($dE/dx$) in the TPC. 
Nevertheless, the system is foreseen to be improved by adding also TOF measurements either in the silicon envelopes around the TPC or in the electromagnetic calorimeter \cite{theildcollaboration2020international}. As a proof of concept, a possible TOF estimator is computed, which uses the first ten calorimeter hits in the ECAL that are closest to the straight line resulting from the extrapolation of the particle’s momentum into the calorimeter, assuming an individual time resolution of 100 ps per hit. Fig. \ref{fig:dEdx_TOF_ILD} shows the complementary between the PID achieved via dE/dx or TOF information at the ILD, making it clear that having both systems would improve the momentum coverage, with TOF being superior in the low momentum regime.

The performance of a PID system based on TOF depends on the time resolution. Ref. \cite{Dharani:414330} shows that smaller time resolutions would boost the PID reach in the low momentum regime at ILD. For example, for a 2 GeV particle, the $\pi/K$ separation power increases from about $5 \sigma$ to almost $17 \sigma$ when the time resolution decreases by one order of magnitude, i.e. from 100 to 10 ps. 
Even with excellent time resolutions though, it must be noted that the momentum range covered via TOF for PID remains limited. As one can see in Ref. \cite{Dharani:414330}, the $\pi/K$ separation drops below $2\sigma$ already at 6 GeV, even in the ideal resolution case scenario. 
To enable physics studies that require PID at high momentum, dedicated PID detectors are needed~\cite{sLoI}.

%\begin{figure}
%    \centering
%    \includegraphics[width=0.5\linewidth]{TOF_ILD.png}
%    \caption{From Ref. \cite{Dharani:414330}. Separation power between kaons and pions as a function of momentum assuming different time resolutions for a time-of-flight system in ILD.}
%    \label{fig:TOF_ILD}
%\end{figure}

Another proposed $e^+e^-$ machine featuring a future multi-TeV collider is CLIC. The CLICdet \cite{arominski2018detector} detector concept is derived from the ILC detector concepts, adapted to the higher energies and background conditions at CLIC. A fully silicon-based tracking system is foreseen. Bunch trains at CLIC occur at a rate of 50 Hz, with each bunch train comprising 312 bunch crossing separated by 0.5 ns. Triggerless readout once per bunch train is foreseen. To ensure low occupancies, e.g. the required maximum pixel size in the vertex detector is 25x25 $\mu m^2$. Similarly to other $e^+e^-$ machines, the timing requirements at CLIC are driven by the levels to which the background degrades the physics performance of the detector. Low occupancies allow for efficient track reconstruction. Accurate track information is used in the subsequent particle-flow reconstruction step, where all charge and neutral particles are reconstructed, including particles from beam-induced background. By requiring a hit-time resolution below 5 ns for the vertex and tracking detectors, and 1 ns resolution for calorimeter hits, all the particle-flow objects have sub-ns time resolution (see Section 2.5 in Ref. \cite{linssen2012physics}). This information is used to remove particles from beam-induced background via momentum and timing cuts as a function of the angular region and of the particle type. As a result, CLIC physics performance is essentially preserved at all foreseen centre-of-mass energies. Dedicated 4D tracking studies with ps-level resolution in some of the tracking layers have not yet been performed in CLIC. It has to be noted though that CLIC assumes to run at 380 GeV and above, thus the impact of fast timing in a TOF system is expected to have less added value for CLIC than for $e^+e^-$ collisions at lower centre-of-mass energies.

Circular $e^+e^-$ colliders offer the possibility of multiple interaction points with plenty of opportunities for innovative and complementary designs. For example, the FCC-ee could host 4 interaction points. At the moment, two concepts have been proposed \cite{FCC_snowmass}: the CLIC-like Detector (CLD) which largely re-uses the CLICdet proposal already discussed above and the International Detector for Electron-positron Accelerator (IDEA) which, rather than a silicon tracker, foresees a drift chamber surrounded by a layer of silicon detector, similarly to ILD at the ILC.

PID capabilities are as important as for linear $e^+e^-$ colliders, considering their impact on the flavor physics program, LLP searches and jet flavor tagging, so even though TOF systems are not yet integrated in the designs, as described in Ref. \cite{FCC_snowmass} the advancement in 4D tracking detectors with good timing resolutions makes TOF detectors an attractive addition to any detector concept at future circular $e^+e^-$ colliders.

%Substituted with the text above to implement Lucie's suggestions
%Two detector concepts are designed based on those studied for the ILC. The timing requirements at CLIC are driven by the levels to which the background degrades the physics performance of the detector. Assuming that the occupancies in the elements of the tracking detectors are sufficiently low that efficient track reconstruction is possible, there is unlikely to be a significant impact on the quality of the reconstructed tracks. Hence the main impact of the background will be on the reconstruction of jets. From detailed studies (see Section 2.5 in Ref. \cite{linssen2012physics}) on the W-boson mass resolution in simulated $W \rightarrow qq$ decays, it was concluded that at CLIC the required hit-time resolution must be below 5 ns in the tracking and vertexing detectors and at most 1 ns in the calorimeters. This hit-time information is used during the reconstruction of particle-flow objects, which starts with track reconstruction by using only tracker hits within a 10 ns time window around the physics event and then combines them with information from calorimeter hits. As a result, all the particle-flow objects (charged and neutral) have sub-ns time resolution and this information can be used to remove particles from beam-induced background via both momentum and timing cuts as a function of the angular region and of the particle type.

\subsection{Muon Collider}

Experiments at muon colliders have huge physics potential. A muon collider as a Higgs factory might allow to directly measure the mass and width of the Higgs boson at highest precision~\cite{ref:mucol:hfac}. On the other hand, muon colliders have a potential to achieve collision energies of tens of TeV with a relatively small size for a collider ring, thus reaching way further than $e^{+}e^{-}$ colliders and having a physics reach on-par with hadron colliders with hundreds of TeV in collision energy~\cite{ref:mucol:highe}. The current timeline to complete the accelerator R\&D and be ready for the construction of a muon collider is estimated to be earliest in the latter part of the 2030s~\cite{ref:mucol:accRM}. 

The major challenge for a muon collider experiment is that muons are unstable particles and naturally decay. The decaying muons within the colliding beams will create, for each beam crossing, a spray of hundreds of million particles entering a muon collider experiment. Out of those, an order of a million particles is charged. This multiplicity of particles entering the detector volume is expected after the muon collider experiment has already been shielded by so-called nozzles in the forward region, blocking the volume of $|\eta| \lesssim 2.5$. The background induced by these particles is commonly referred to as the beam-induced background (BIB)~\cite{ref:mucol:bib}. The BIB primarily consists of low energy photons and neutrons with a small fraction of charged hadrons, muons and electrons also present. 

The presence of the BIB puts stringent requirements on a tracking detector. Firstly, the high number of particles entering the detector region leads to high levels of radiation and thus detectors need to be radiation hard, similarly to the detectors at hadron colliders. Secondly, hits produced by the BIB particles complicate data readout and make track reconstruction at the muon collider a very challenging task. 

Yet, as the decay of muons is a stochastic process, several advantageous design aspects for tracking detectors can be thought of to suppress the impact of the BIB. Most of the BIB particles enter the detector from the two forward regions and do not originate in the collision area. Precise timing of detector hits would be able to reduce the BIB by a large fraction as we only need to consider hits consistent with the collision time. Furthermore, if timing of the hits can be correlated among adjacent layers of a tracking detector, a further filtering can be done by only considering hits consistent with being produced by the same particle. Initial studies indicate that single hit resolutions of 20-30 ps are sufficient to reduce the BIB to a manageable level. Additional suppression can be achieved if the tracker can obtain directional information, as for example is being done with the $p_{\mathrm{T}}$ modules of the CMS outer tracker upgrade for the HL-LHC~\cite{ref:mucol:CMSupgrade}. Besides the requirements of precise timing and directionality, also a high spatial resolution is needed to achieve a low detector occupancy. Simulation studies show that small pixels at a size of about $(25\,\mathrm{{\mu}m})^2$ are needed at the innermost layers of a tracking detector while even at the outermost layers strips with length of at most few cm are required~\cite{ref:mucol:det}.

These requirements on small pixels/strips with precise timing and directional information will allow to not only handle the BIB, but will also enable a muon collider experiment to take high quality data for precision measurements and searches for new physics at highest energies~\cite{ref:mucol:perfm}. Therefore, there is a high need of R\&D efforts towards 4D and 4D with additional directional information (5D) tracking within the muon collider community.

\subsection{Electron Ion Collider} 

The Electron-Ion Collider (EIC)~\cite{AbdulKhalek:2021gbh} is a new accelerator facility to be built at Brookhaven National Laboratory in the United States. In 2031, the machine will begin colliding high-energy electron beams with high-energy proton and ion beams to study the spatial and spin structure of nucleons and nuclei. Due to the small $ep$ and $eA$ cross-sections, the collision rate at the EIC will be 500 kHz or less with a total particle production rate of about 4 million per second. Therefore, the requirement on the radiation tolerance and occupancy of the detectors at the EIC are considerably relaxed compared to those at the hadron colliders. 

Precision timing information would provide two powerful capabilities to the EIC experiments: particle identification based on time of flight, and vertex identification for far-forward hadrons resulting from exclusive interactions. These goals could be achieved with 4D detectors with time resolutions on the order of \SI{30}{\pico \second}, and spatial resolutions of \SIrange{15}{150}{\micro\m}, depending on the location.

Timing detectors based on AC-LGADs, which will be described in Section~\ref{ssec:lgads}, are envisioned to provide timing capability and are included in all of the submitted EIC detector proposals. These would serve as the first demonstration of truly 4-dimensional tracking layers in a collider experiment, building on the foundation of the CMS and ATLAS timing layers. The performance requirements impose a great challenge for the AC-LGAD sensors, front-end readout ASIC, off-detector electronics, as well as the mechanical and cooling system design. A R\&D project has been established to develop a common approach for these detectors so that they can share the same design to the extent possible.

\section{Time resolution}
Time resolution is the crucial new ingredient to achieve 4D tracking. In this section a brief introduction to its separate components is discussed. The time resolution of a detector can be expressed as follows:
\begin{equation}
\label{eq:timeres}
    \sigma^2_{timing} = \sigma^2_{timewalk} + \sigma^2_{Landau} + \sigma^2_{Jitter} + \sigma^2_{TDC}
\end{equation}
The first two components, time walk and Landau, are intrinsic to the sensor. 
Time walk refers to the variation of the deposited charge event-by-event, shifting the threshold crossing earlier for larger charges and later for smaller charges. 
This component can be minimized by either using a variable threshold or a corrected constant threshold.
A variable threshold is, for example, the Constant Fraction Discriminator (CFD) where the 20-50\% of the pulse maximum is used as the time of arrival.
A more common method in integrated electronic chips is to correct the time walk on the time of arrival (TOA) with a calibration based on the time over threshold (TOT), which serves as a proxy for the charge.
Beyond variation in the total ionization, the Landau term represents the spatial variation of the deposited charge along the path of a minimum ionizing particle (MIP), as charges from different depths are collected at different times. Since a MIP usually traverses the entire sensor perpendicularly, this component is smaller for thinner devices or sensors with short drift times.
In devices where the S/N is high and time walk can be adequately corrected, the irreducible Landau component is usually the limit of the achievable time resolution.

The second two components are related to the readout chip's electronics. The jitter is described as
\begin{equation}
\label{eq:jitter}
    \sigma^2_{Jitter} = \frac{Noise}{dV/dt} \approx \frac{T_{rise}}{S/N}
\end{equation}
Therefore $\sigma^2_{Jitter}$ is proportional to the rise time and inversely proportional to the S/N ratio. Since the rise time is proportional to the drift time of charge carriers in the sensor, again a sensor with short drift time and high signal to noise ratio is advantageous.

The time to digital conversion (TDC) component is related to the readout chip's TDC precision to measure TOA and TOT, given by the quantization bin size. In the case of the planned timing detectors at the HL-LHC, they range between 20ps (ATLAS) and 30ps (CMS) for the TOA TDC, and 40ps (ATLAS)  and 100ps (CMS) for the TOT used for the time walk correction. Given that quantization errors are described by an uniform probability density, $\sigma^2_{TDC}$ is given by the bin size divided by $\sqrt{12}$ which is about 5ps.

To summarize, a candidate timing device would need to have the following characteristics: short drift time (reduces rise time), high signal to noise (reduces jitter component), limited thickness in the path of a MIP (reduces Landau component), and small TDC bin size (reduces TDC component). 
These properties can be achieved by exploiting several technologies introduced in the following section.

\section{Sensor technologies}

From the overview of trackers in the diverse future collider detectors considered in Section 1, a common set of requirements emerge. Future tracking sensors will require simultaneous fine resolutions in both time and space of order  5--30 \si{\pico\s} and 5--25 \si{\micro \m}. In some collider scenarios, there are also stringent constraints on the sensor radiation hardness, occupancy, and material budget. And for certain applications, additional features like single-layer directional measurements can be of great value.

Several approaches are being pursued to develop sensors to meet these requirements. One category takes Low Gain Avalanche Detectors (LGADs) developed for precision timing as a starting point, and adapts them to achieve the fine segmentation necessary for tracking. These advanced LGAD concepts are described in Section \ref{ssec:lgads}. 
New materials are also being pursued for LGAD fabrication such as Silicon Carbide, a detailed description of the technology can be found in Ref. ~\cite{2203.08554}.
Alternatively, tracking sensors can be modified to facilitate precision timing through a variety of techniques. These include adopting 3D geometries with short drift times (Section \ref{ssec:3D}); or through closer integration with electronics, for example via induced current detectors as described in Section \ref{ssec:induced} or via monolithic pixel sensors. 

For the latter, excellent spatial resolution can already be achieved along with a reduction of the material budget typical of the monolithic technology compared to the hybrid one. Studies are actively ongoing to obtain radiation hard CMOS sensors with fast timing capabilities. The state-of-the-art results from MALTA \cite{MALTA_talk} show that $<$ 2 ns resolution is already achievable with a full scale working prototype, making this approaches suitable for example at $e^+e^-$ colliders. The FASTPIX \cite{FASTPIX} team proved that, with additional dedicated optimisations, demonstrators can achieve a time resolution of about 120 ps. An investigation of timing in MAPS \cite{MAPS} is also being performed.

\subsection{Advanced LGAD designs}
\label{ssec:lgads}

The most prominent class of 4D tracking sensors proposed for future detectors are based on an evolution from Low-Gain Avalanche Detectors developed for the HL-LHC. LGADs are thin silicon sensors with modest intrinsic gain (5-50) provided by a moderately doped p+ multiplication layer, and achieve \SI{30}{\pico\s} or better time resolution. 
The exceptional time resolution of LGADs comes from the thin bulk (usually 50~$\mu$m or less) that minimizes the Landau component of the time resolution, and high S/N (thanks to the internal gain mechanism) that, together with the short rise time, minimizes the jitter component of the time resolution.
Standard LGADs, however, feature \si{\milli\m} scale pads and rely on Junction Termination Extensions to interrupt the gain layer between channels, which introduces inactive regions. As a result, LGADs cannot be simply miniaturized to a pitch appropriate for tracking.
%J: also mention the ~good radiation hardness of LGADs?
Several sensor concepts have been proposed and demonstrated to make LGAD sensors suitable for tracking, in addition to providing excellent timing resolution. In the following sections, several innovative LGAD designs to achieve high granularity will be introduced.

Current state of the art LGADs can tolerate neutron fluences only up to roughly 1--2~$\times10^{15}$~MeV neq~$\si{\cm^{-2}}$, due to the loss of gain implant acceptors. To be used in future hadron or muon collider trackers, LGADs must be developed with radiation tolerance an order of magnitude larger. 

%Many of these fall in the category of AC-LGADs, which feature continuous gain layer, a resistive n+ surface layer, and AC-coupled readout electrodes. Because the gain layer is uninterrupted, AC-LGADs  achieve a fill factor of $100\%$, and electrodes can be designed with smaller pitch and size than standard LGADs. A key feature of AC-LGADs is the signal sharing between electrodes, which can be used to obtain position measurements with resolution much smaller than $\mathrm{(bin~size)}/\sqrt{12}$. One major advantage of AC-LGADs is that they can obtain the spatial resolution necessary for future colliders with a coarser pitch and hence a reduced number of readout channels. 

%Recent test beam measurements have demonstrated that AC-LGADs can achieve simultaneous \SI{30}{\pico \s} and  \SI{5}{\micro \m} resolution with strips of pitch 100-200 \si{\micro\m}~\cite{RD50ACLGAD}. This spatial resolution  represents a factor of 5-10 improvement beyond what would be obtained in binary readout, thanks to the sharing of signals between adjacent channels. Other studies have demonstrated similar performance \cite{TORNAGO2021165319}\cite{Apresyan_2020}.

\subsubsection{AC-coupled LGADs}

The most advanced high granularity LGAD design realized so far is the AC-coupled LGAD (AC-LGAD).
In contrast to standard LGADs, AC-LGADs feature a continuous gain layer and a more resistive n+ surface layer. The signal is capacitively (AC-) coupled over the equally continuous dielectric to the metal readout electrodes. Because the gain layer is uninterrupted, AC-LGADs  achieve a fill factor of $100\%$, and electrodes can be designed with smaller pitch and size than standard LGADs. A key feature of AC-LGADs, brought on by the common gain and n+ layer,  is the signal sharing between electrodes, which can be used to obtain position measurements with resolution much smaller than $\mathrm{(bin~size)}/\sqrt{12}$. Hence, one major advantage of AC-LGADs is that they can obtain the spatial resolution necessary for future colliders with a coarser pitch and thus a reduced number of readout channels. 

Recent test beam measurements have demonstrated that AC-LGADs can achieve simultaneous \SI{30}{\pico \s} and \SI{5}{\micro \m} resolution with strips of pitch 100-200 \si{\micro\m}~\cite{FNALtb2021}. This spatial resolution  represents a factor of 5-10 improvement beyond what would be obtained in binary readout, thanks to the sharing of signals between adjacent channels. Other studies have demonstrated similar performance \cite{Tornago:2020otn}\cite{Apresyan_2020}.

\begin{figure}[htp]
    \centering
    \includegraphics[width=0.26\textwidth]{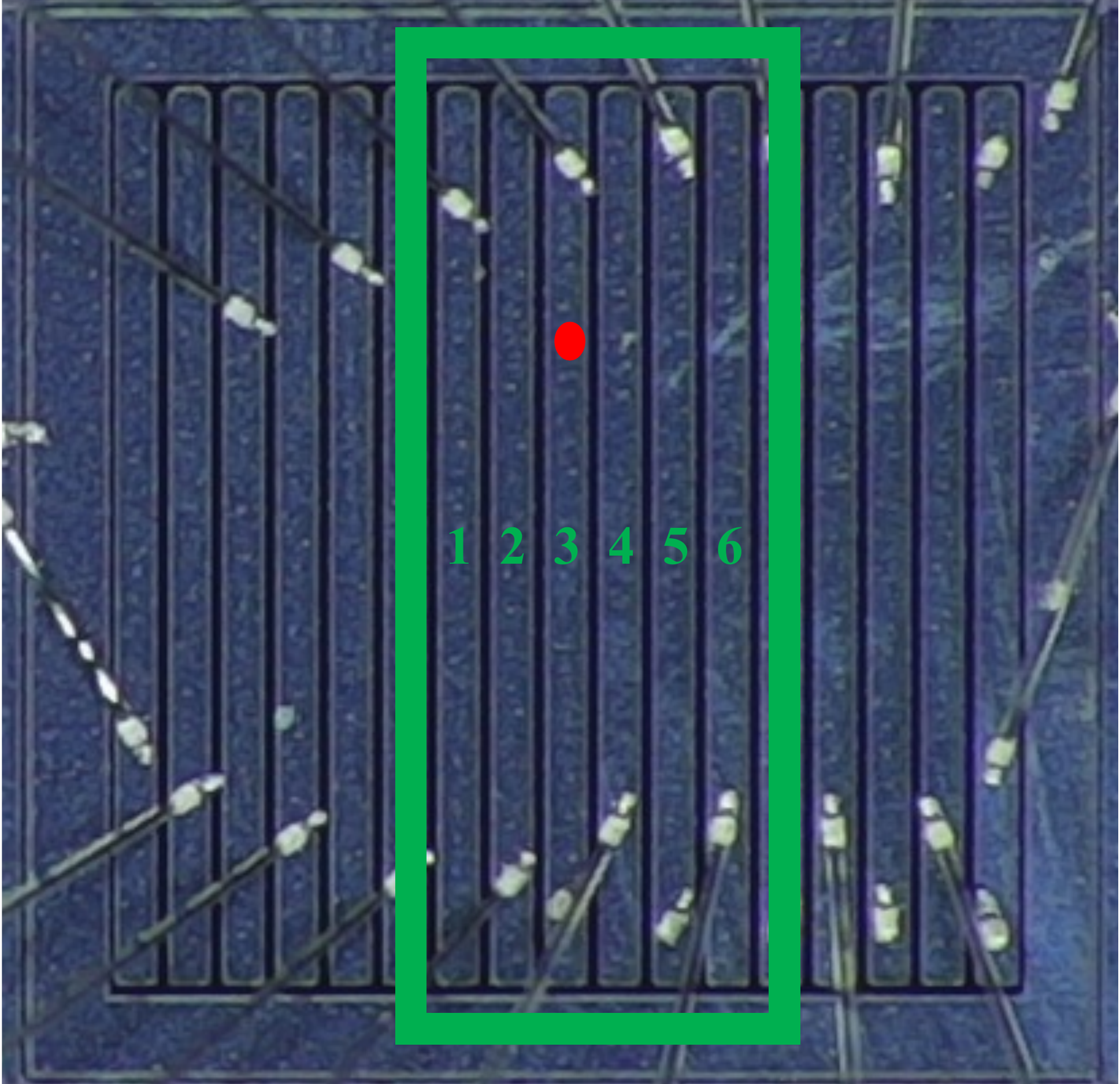}
    \includegraphics[width=0.33\textwidth]{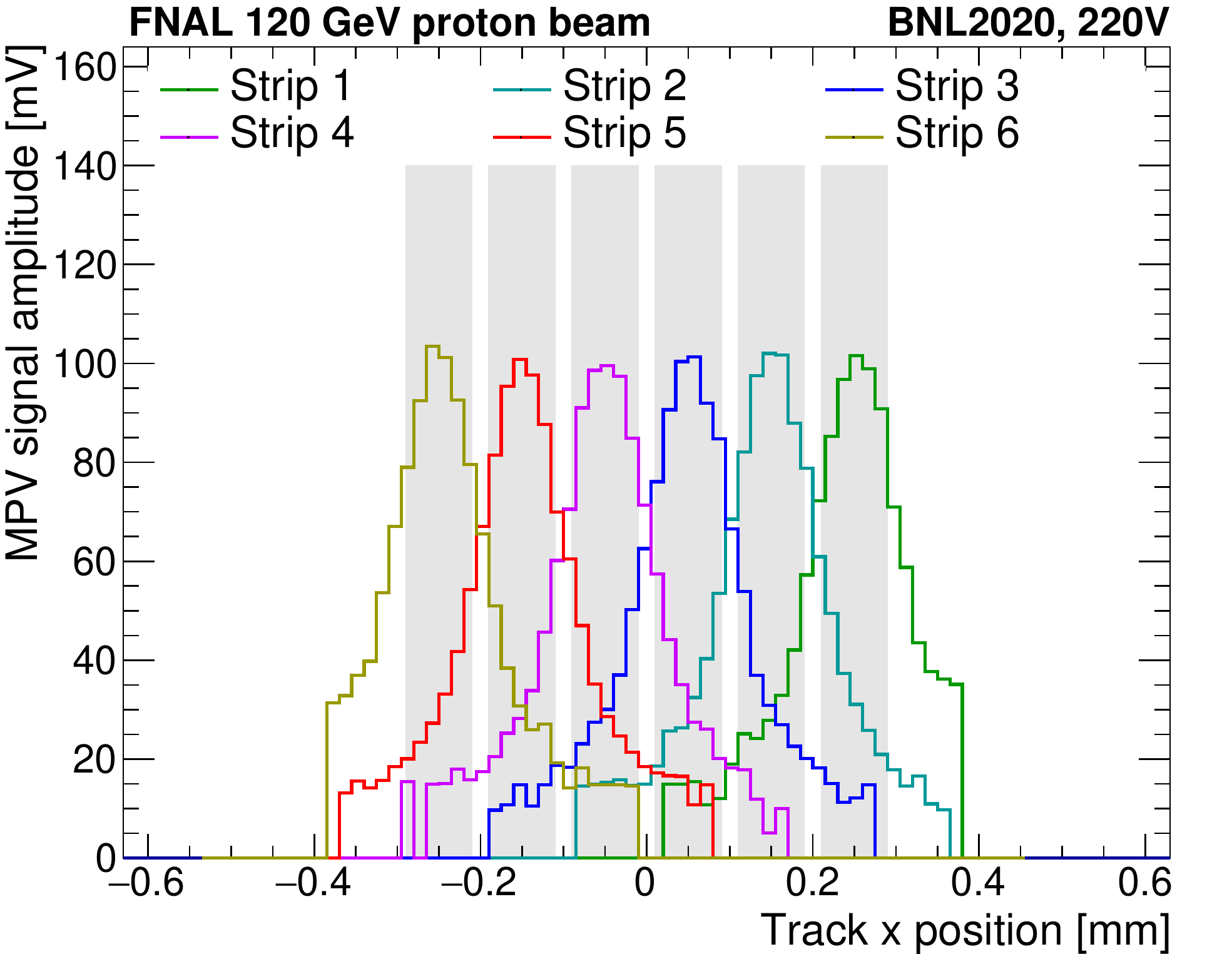}
    \includegraphics[width=0.33\textwidth]{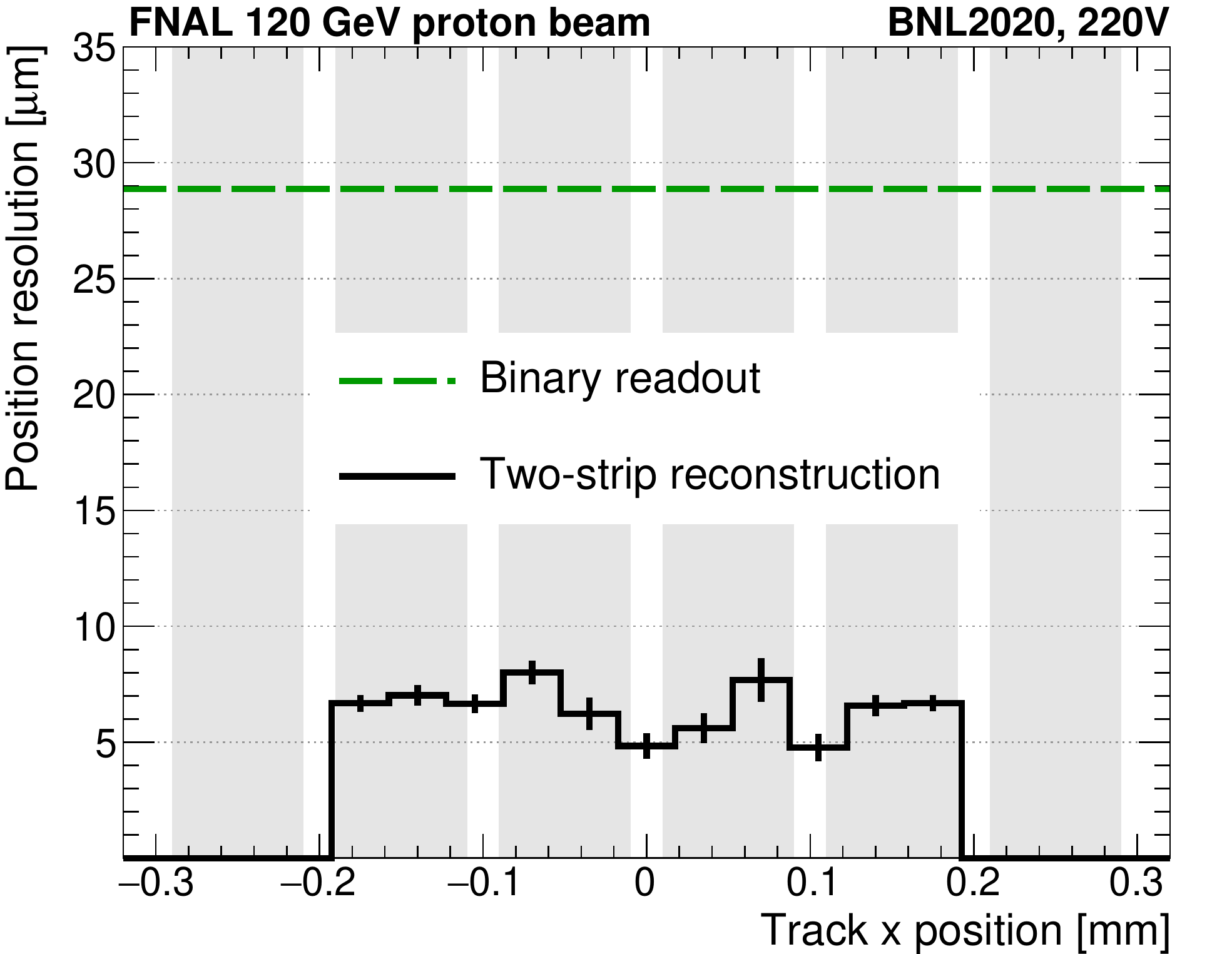}
    \caption{
    AC-LGAD strip sensor prototype produced at Brookhaven National Laboratory, with \SI{100}{\micro\m} pitch. The green box indicates readout channels (left).
    Signal amplitudes shared between various channels as a function of the particle impact parameter at the Fermilab Test Beam Facility (center).
    Spatial resolution as a function of proton impact parameter, including roughly \SI{5}{\micro\m} contribution from the reference tracker (right). The grey areas indicates the metallized regions on the sensor surface.}
    \label{fig:AC_LGAD_testbeam}
\end{figure}

AC-LGADs have several parameters that can be tuned to optimize the sensor response to the specific application.
The geometry of the electrodes in terms of pitch and pad dimension is the most important one, however also the n+ sheet resistance and the thickness of the dielectric between n+ and metal electrodes influence the charge sharing mechanism, as well as signal characteristics such as the undershoot (recovery to baseline) after a signal.
All of the AC-LGAD parameters affect the final performance of the sensor, and therefore it is necessary to balance these properties to achieve the best timing and spatial resolution with respect to the channel density and occupancy.

For optimizing the sensor design in terms of the abovementioned parameters, simulation with TCAD software, such as Silvaco or Synopsys Sentaurus, is an important tool. On the other hand, experimental data from prototype sensors serves as a vital input to simulation models in order for them to provide a good representation of the observed sensor performance, and have predictive power for future devices. 

As neither the gain layer, n+ layer nor dielectric have significant underlying structure, it is possible to arrange the metal of the AC-LGAD readout electrodes in any desired shape and size. This allows to optimize the electrode geometry to tune the charge sharing to the specific application. 
For example, circular or cross-shaped metal electrodes, instead of squares, could further improve position resolution through the utilization of advanced reconstruction algorithms. Electrodes can also be arranged on a triangular grid, instead of square or rectangular alignment, to further enhance charge sharing. Furthermore, electrodes can be shaped to have different charge sharing in the X and Y direction (e.g. micro-strips) to optimize the channel density to the sensor resolution in both directions.
A recent sensor production by FBK allows systematic evaluation of various electrode patterns and geometries\cite{RD50_RSD2}.
Another proof of principle to obtain different metal electrode geometries, featuring a modification process of metal electrodes executed at BNL, was made~\cite{BNLetch}. The procedure was successful, showing that the top metal geometry of AC-LGADs can be modified. Thus, a universal metal design could be produced and subsequently etched to meet the requirements of any specific application.

\subsubsection{Buried Layer LGADs}

Irradiation campaigns have indicated that LGADs with deep and narrow gain layers are more radiation hard than LGADs featuring shallower and broader gain layers. Gain layers are generally obtained by means of ion implantations which can implant boron at a maximum depth of about 2 $\mu$m using very high and not easily available implantation energies. Furthermore,  deeper implants generally introduce more broadening than shallower implants. A way to circumvent the problem, at the expense of a complication in the process, is to implant the boron layer at low energy and then to bury it under a few microns of epitaxially grown silicon, obtaining in this way a deep and narrow gain layer (if high thermal cycles are avoided in the subsequent process).  The method can be applied to either standard DC-coupled LGADs or AC-LGADs to improve the radiation tolerance of the gain implant.
A first fabrication has been completed but suffered from a high leakage current due to a poor epitaxial deposition. Another fabrication is on-going.

\subsubsection{Trench-isolated LGADs}

\begin{figure}
    \centering
    \includegraphics[width = 0.8 \textwidth]{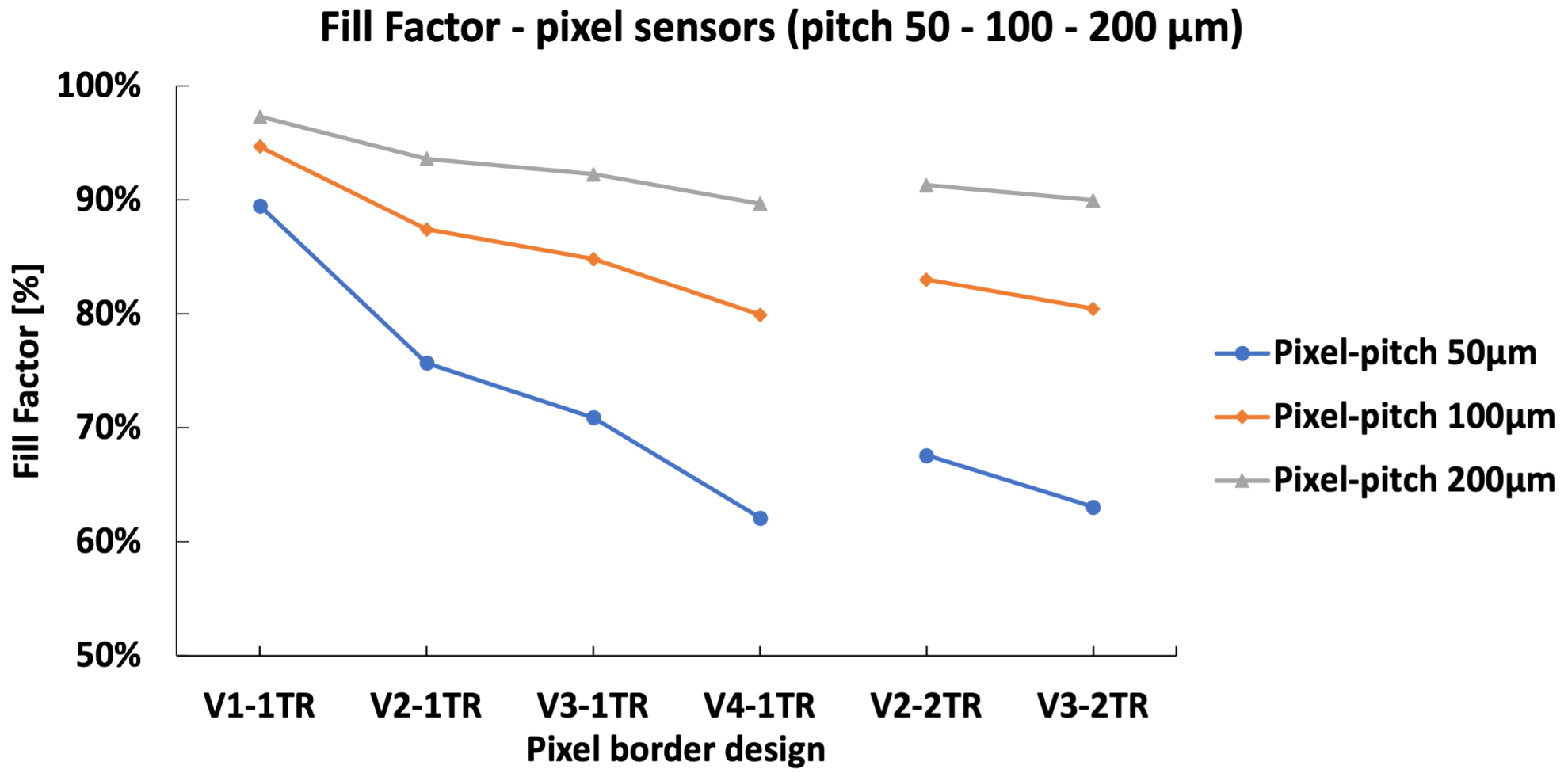}
    \caption{Expected fill factor of square TI-LGAD pixels for different channel border layouts. The width of no-gain area was measured on TI-LGAD pad sensors using a pulsed laser~\cite{FerreroTredi2022}.}
    \label{fig:extraFF}
\end{figure}

Trench-isolated (TI) LGADs aim to to achieve segmentation with high fill-factor by substituting the structures used for the termination of the gain layer (such as junction termination extension) and channel isolation with narrow trenches filled with dielectric material.
The width of the trenches is about 1~$\mu$m, allowing for a substantial reduction of the distance between gain layers of neighboring channels with respect to standard LGADs~\cite{9081916}.
A first TI-LGAD production at FBK demonstrated the isolation between pixels using trenches and showed the possibility to achieve no-gain areas between channels a few microns wide~\cite{9081916,PaternosterNIMA2021}.
A second production was realized at FBK for the CERN RD50 collaboration\footnote{RD50 - Radiation hard semiconductor devices for very high luminosity colliders \url{cern.ch/rd50}}.
This production allowed for more systematic studies of TI LGADs, exploring different fabrication parameters and channel border designs.
A first characterization of this production shown that the trenches do not result in a premature breakdown of the sensor, and the current-voltage characteristic of TI-LGADs is dominated by the effect of the gain layer for many of the fabrication parameters combinations employed in this batch~\cite{BishtPicosecond2021}.
Measurements using pulsed lasers shown no-gain areas between channels ranging from 3 to 15 $\mu$m depending on the channel border layout and fabrication process, while maintaining the gain of standard LGADs~\cite{BishtPicosecond2021,SengerVCI2022,FerreroTredi2022}.
Similar results were obtained using TI-LGAD strip sensors~\cite{MazzaPrivate2022}.
The expected fill factor for square TI-LGAD pixels of different pitch is shown in figure~\ref{fig:extraFF}.
The no-gain area was measured for different border layouts on pad sensors using a pulsed laser.
The time resolution expected from LGADs was also demonstrated on TI-LGAD pad sensors using electrons from a $^{90}$Sr source~\cite{FerreroTredi2022}.
The radiation hardness of the isolation provided by trenches was tested up to a fluence of $3.5 \cdot 10^{15}$~n$_{eq}$cm$^{-2}$ using reactor neutrons~\cite{SengerVCI2022} and up to a x-ray dose of 10~Mrad~\cite{FerreroTredi2022}.
These tests show that the isolation between channels is maintained for all the tested fluences and doses.
The width of the no-gain area is maintained or reduced in irradiated sensors, however the latter case could be a result of the deactivation of the gain layer with irradiation.
The gain layer employed in the sensors used in these radiation studies was not optimized for radiation hardness and lacked carbon coimplantation, therefore the results obtained in terms of gain and time resolution as a function of fluence are not expected to match more radiation hard designs.
The use of carbon coimplantation and more radiation hard gain layers is expected to be compatible with the TI-LGAD technology.
TI-LGADs may have an advantage with respect to AC-coupled LGADs for applications with high occupancy constraints, since the signals are typically contained in a single channel, rather than dispersing in multi-channel clusters as in AC-coupled sensors.

%An alternate way to achieve segmentation with high fill-factor is to use narrow trenches, filled with insulator to separate the pixels (or strips). The deployment of such trenches all around the electrodes allows to get rid of the JTE, to bring the gain layer closer to the edge of the n-plus implants, and  to bring the pixels close together. Dead areas of a few microns have been demonstrated in the first FBK prototype productions~\cite{9081916}.

%Trench-insulated LGADs may have an advantage with respect to AC-coupled LGADs for applications with high occupancy constraints, since the signals are typically contained in a single channel, rather than dispersing in multi-channel clusters as in AC-coupled sensors.

\subsubsection{Double Sided LGADs for 5D Tracking}

The Double Sided LGAD (DS-LGAD) adds a readout layer to the p-side of the LGAD structure. This allows for double-sided readout with the p side reading out the slower-drifting holes. For a device with the bulk thickness large compared to the pixel pitch, the p-side readout can function as a miniature time projection chamber with the 
drift time providing information on the depth of origin of the charge cloud. The signal p-side collects two components, 
holes from the primary ionization followed by the larger number of holes generated at the gain layer. This provides 
a unique signature of the pattern of charge deposit within the device that could enable measurement of the track angle as well. The fast rise signal can be read by a large area cathode, limiting the number of complex, high-power fast amplifiers and digitizers. The p-side provides a large, slower signals that can be read out with electronics with lower complexity and power consumption.
%The overall features of the DS-LGAD are described in %\cite{Lipton:2021DSLGAD}.

The characteristics of the thick ($> 5\times pitch$) DS-LGAD are sensitive to the interplay between the device thickness, gain layer depth and doping, and the applied voltage. In a buried layer device the depth and doping of the gain layer set the operating point and the drift field. These can be tuned to achieve the required characteristics. Diffusion width and time of arrival of the holes from gain layer amplification could then be used to provide excellent position and good track angle resolution. Since DS-LGADs would be thicker than conventional LGADs, there would be a tradeoff in time resolution in order to achieve sensitivty to the track angle.

\subsubsection{Deep-Junction LGADs}
The Deep-Junction LGAD (DJ-LGAD)~\cite{Ayyoub:2021dgk} is a new approach to the application of controlled impact-ionization gain within a silicon diode sensor. The term ``deep-junction" arises from the use of a p-n semiconductor junction buried several microns below the surface of the device. 
The buried junction is formed by abutting thin, highly-doped p+ and n+ layers, with the doping density chosen to create electric fields large enough to generate impact ionization gain in the narrow buried junction region. 
Additionally, the doping densities chosen for the p+ and n+ layers are balanced so that when the sensor is fully depleted, the electric field outside of the junction region, while large enough to saturate the carrier drift velocity, is significantly less than that require to create impact ionization gain. 
This preserves the electrostatic stability at the segmented surface of the detector, thus in principle permitting the production of DC-coupled LGADs with arbitrarily fine granularity. 
It can be pixelated like a standard silicon detector (\textless~50~um) with perfect insulation between pads since no JTE structure is required between pixels. 
%No JTE structure is required for a DJ-LGAD array, as the buried gain layer ensures a uniform gain performance across channels.
The DJ-LGAD approach is seen to hold significant promise towards the development of a highly-pixelated DC-coupled silicon diode sensor with substantial internal gain and precise temporal resolution.
The effect of radiation damage on DJ-LGAD needs to be evaluated, the lightly doped n region after the buried gain layer will be affected starting from a low fluence.

\subsubsection{Thin LGADs}
Utilizing thinner sensors may provide a path to extend the radiation hardness of silicon sensors. \cite{RADTREDI, RADCPAD}.
In recent years, a saturation of the charge trapping effect in silicon was observed \cite{RADSAT}.
However, despite the saturation, at a fluence of $10^{17}-10^{18}$~Neq a standard 300~um silicon detector would still need several thousand volts to deplete.
In contrast, a thin sensor can reach full depletion even at very high fluence at much lower voltages: 500~V of full depletion for a 50~um sensor at $10^{17}$~Neq.
The collected charge for thin sensors, however, would be too small to be efficiently detected by readout electronics.
Thanks to the intrinsic charge multiplication thin LGADs can be used for this application.
It has been shown that sufficient gain is present for time resolution measurements purposes only until a few $10^{15}$~Neq, but less gain is needed for regular hit detection.
Furthermore at high fluences gain in the bulk ``p'' region of the sensor can be activated by increasing the bias voltage applied to the sensor.
These statements gives an indication of suitability of thin LGADs even at extreme fluences for hit detection, such as the tracking system very close to the interaction point in future hadron colliders.

\subsection{3D silicon sensors}
\label{ssec:3D}

Since their introduction in 1997~\cite{PARKER1997328}, sensors with three-dimensional electrodes (3D silicon sensors) have been widely consolidated, being presently used in LHC experiments (CMS-PPS~\cite{Obertino:2020ovd}, ATLAS-IBL~\cite{ATLASIBL:2018gqd}). Such sensors are characterised by cylindrical electrodes, penetrating the bulk material, perpendicularly to the surface. 3D sensors with cylindrical electrodes have shown to maintain high performance up to fluence close to $10^{17}$ 1~MeV neq/cm$^2$~\cite{Lange_2018,MANNA2020164458}. The specific structure of 3D sensors decouples the charge carrier drift distance from the sensor thickness. This allows reduced inter-electrode spacing, thus minimising the probability of charge trapping by defects generated by radiation. The geometry of 3D sensors is also very beneficial in terms of timing performance. Very short collection times can be achieved without reducing the substrate thickness, thus preserving
the signal amplitude. The fact that the charge carriers are collected perpendicularly to the
sensor thickness minimises time uncertainties due to nonuniform ionisation density, delta rays
and charge carrier diffusion. Moreover, the vertical geometry of the electrodes allows a large
margin of choice in defining the pixel layout and the structure of the sensitive volume. This
property can be exploited in minimising unevenness in the electric field and increasing the
uniformity in signal response to obtain very high time resolutions. 
In the last years, results have been obtained by the TimeSPOT collaboration \cite{Anderlini_2020,Brundu:2021hed}, which show that by proper design of the sensor layout and electrode geometry, the time resolution can be pushed well below 20~ps at room temperature. The measured resolution is limited by the front-end electronics performance, while the sensor intrinsic resolution is estimated around 10~ps. TimeSPOT sensors have been designed and fabricated in a so-called 3D-trench geometry (linear parallel trenches for bias and collection electrodes). The sensitive pixel volume size is 55x55x150~$\mu m^3$ and is operated at a bias voltage around 100~V. 
A general feature of 3D sensors are the dead areas represented by the biasing and collecting electrodes, which introduce a detection inefficiency for particles traveling perpendicular to the sensor surface. This effect has been studied in detail for 3D silicon sensors based on columnar geometry, which are used in the ATLAS IBL~\cite{ATLASIBL:2012zmo}. Test beam studies have been recently performed also on 3D-trench TimeSPOT pixels. Preliminary results show that an efficiency of almost 100\% is reached for tracks with 20 degrees inclination, while maintaining the same time resolution.
Dedicated integrated CMOS 28~nm read-out electronics is under development. A small prototype ASIC, named Timespot1, featuring a matrix of 1024 pixels and integrating one TDC per pixel channel has been recently tested being capable of a time resolution around 30 ps on the full read-out chain~\cite{Piccolo:2022hsz}.

%\subsection{MAPS}
%\label{ssec:maps}
%Fine spatial resolution and course time resolution: CMOS MAPS. Coordinate with CMOS paper

%Valentina: for Malta, there is public material here we can use as a reference (including a section on timing)
%https://twiki.cern.ch/twiki/bin/viewauth/Atlas/MaltaApprovedPlots

%and also this CERN EP RD seminar:
%https://indico.cern.ch/event/1074066/#2-measurement-results-from-fas

\subsection{Induced Current Detectors}
\label{ssec:induced}
An induced current detector uses the same sensor as a traditional silicon detector but utilizes small pixel pitch and 3D integration (3DIC) techniques to create a low-capacitance pixel unit cell and readout chain. This not only limits the amount of noise in the system but also enables the detection of the induced current as described by the Shockley-Ramo theorem~\cite{Shockley:1938itm, Ramo:1939vr}. The Shockley-Ramo current is the current induced at the readout electrode from mobile charge carriers within the sensor. It has a very fast rising edge $(\sim15~\textrm{ps})$~\cite{Lipton:2019drv} and can be used to precisely timestamp track hits. The Shockley-Ramo current is dependent on the weighting field within the sensor and the depth of charge deposition. As a direct result, it has a complicated bi-polar signal shape that integrates to zero over the course of several nanoseconds. This signal combined with the drift current creates a pulse shape that is dependent on the charged particle’s angle of incidence.

A detector that is sensitive to the effects of the Shockley-Ramo current has two critical features: precise $(\sim15~\textrm{ps})$ time resolution and angle of incidence information~\cite{RSCPAD}.  The angular information could be used to quickly identify particles with large transverse momentum (\pt) for an L1 trigger, as high-\pt particles have high angles of incidence. Additionally, it could be used to greatly decrease the complexity of track reconstruction by reducing the number of hits that must be considered when track seeds are generated. These capabilities make it a very attractive detector technology for future collider experiments and requires the continued investment in advanced ASIC design and 3DIC techniques.

\begin{figure}[htp]
    \centering
    \includegraphics[width=0.9\textwidth]{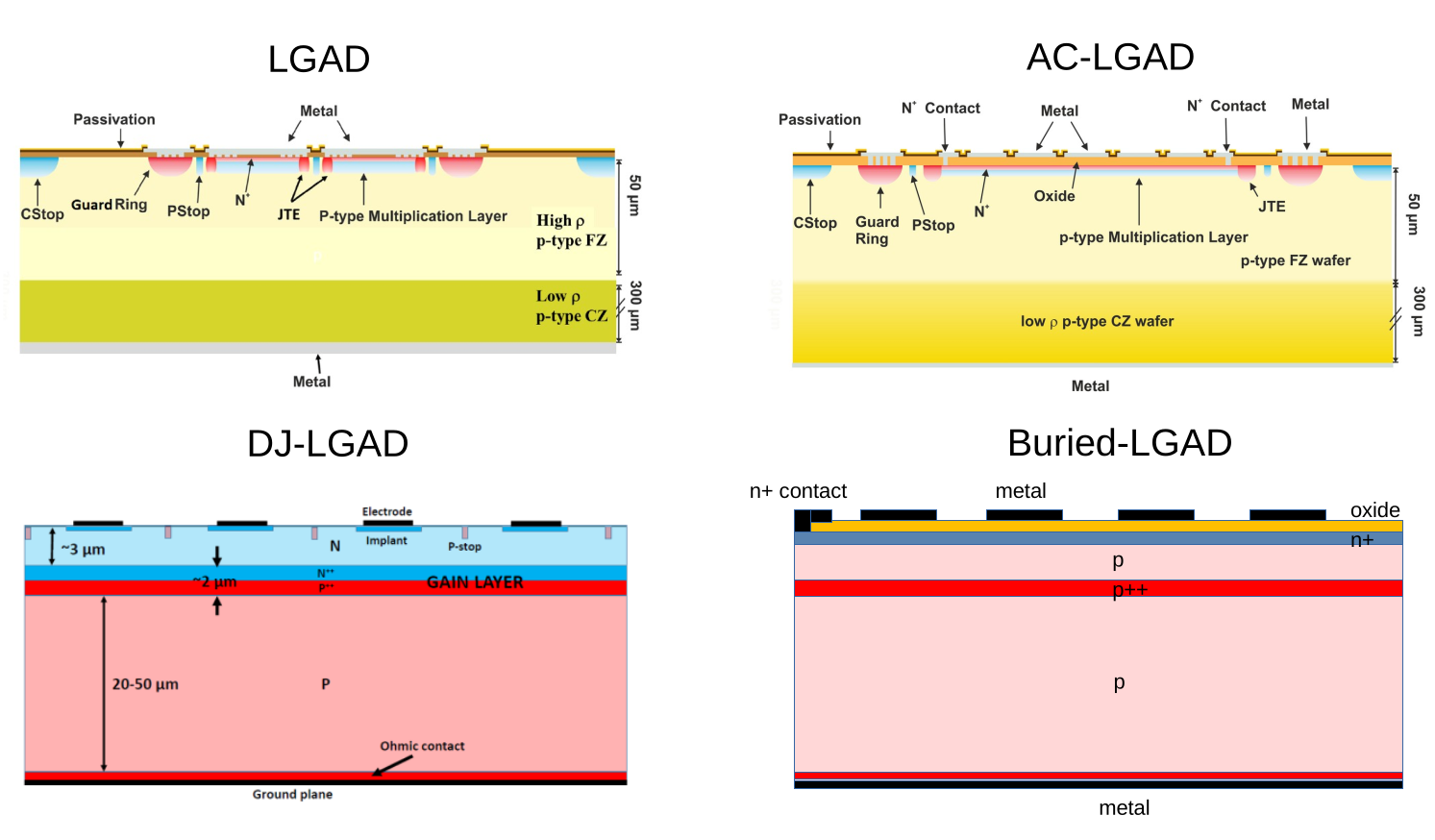}\\
    \includegraphics[width=0.45\textwidth]{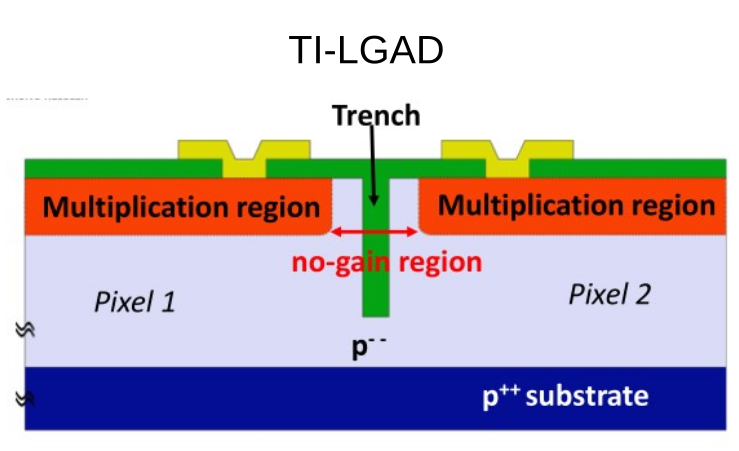}
    \includegraphics[width=0.4\textwidth]{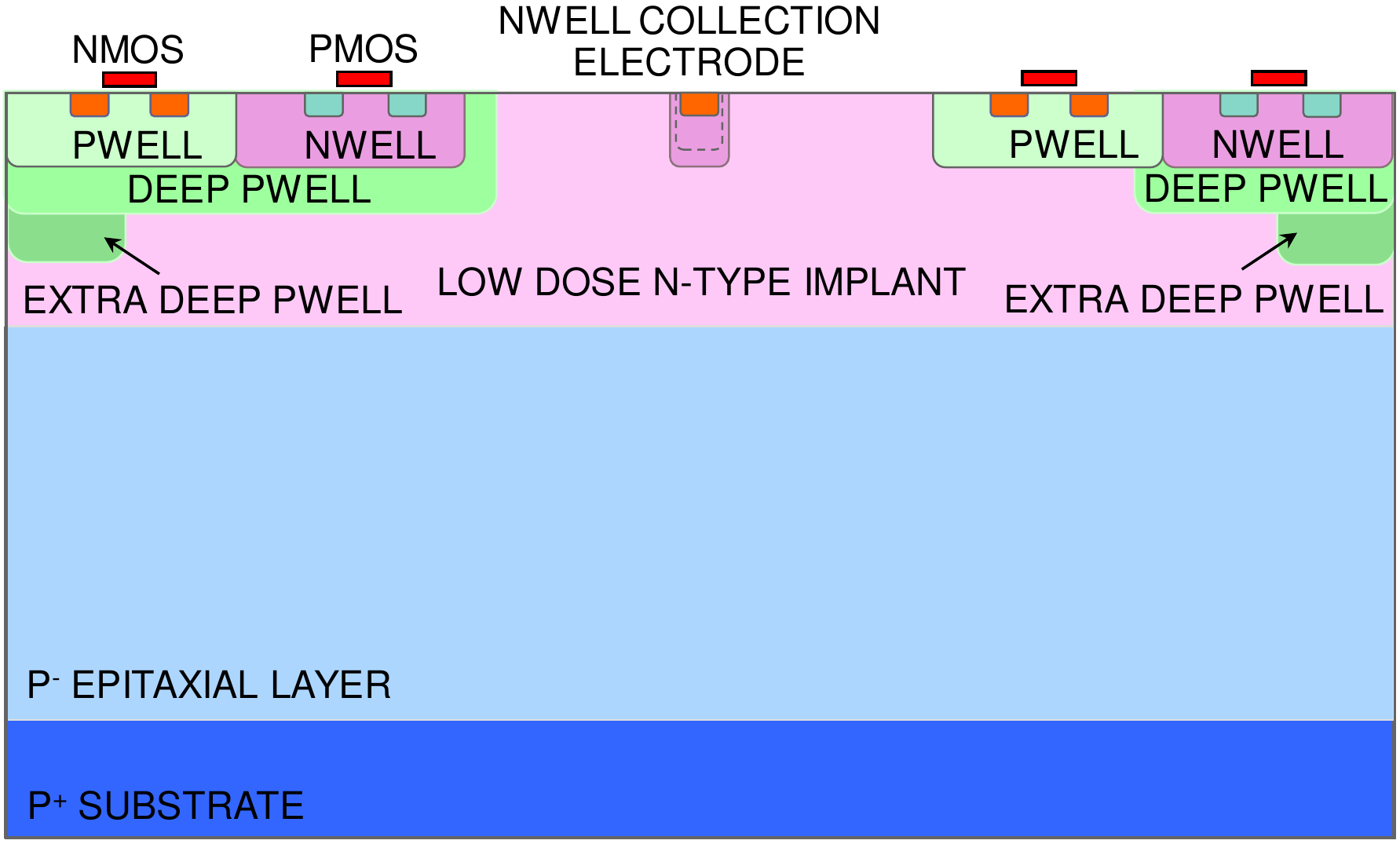}\\
    \includegraphics[width=0.4\textwidth]{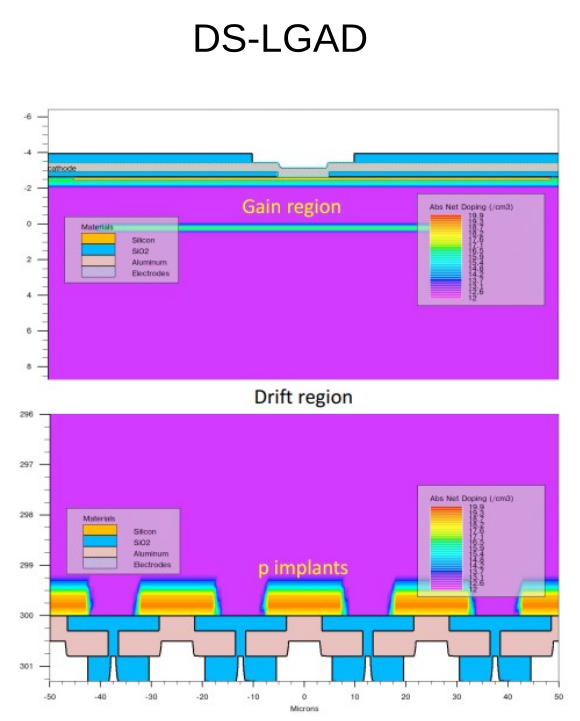}
    \includegraphics[width=0.4\textwidth]{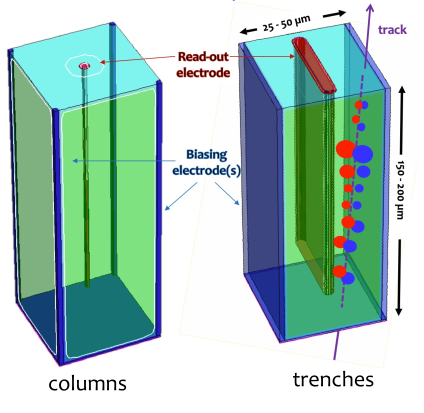}
    \caption{
    Schematic for different types of LGADs. From left to right and top to bottom: standard LGAD, AC-LGAD, DJ-LGAD, Buried-LGAD, TI-LGAD, Malta HV-CMOS \cite{MALTA}, DS-LGAD, 3D normal and trench silicon sensor.}
    \label{fig:sensor_types}
\end{figure}

\subsection{Comparison of sensor technologies}

The diverse sensor designs introduced in this section feature unique advantages and weaknesses, and the choice of technology must be tuned to the specific application. The schematics of the various sensor designs are summarized in Figure~\ref{fig:sensor_types}.

The constraints of a particular experiment combined with the tradeoffs inherent to each sensor drive the appropriate sensor choice.
For example, AC-LGADs have a very high fill factor, around 100\% of the active area, and rely on charge sharing for hit reconstruction: this allows for high position resolution for a sparse readout in low occupancy environments. 
The low channel density reduces the power consumption density which is one of the main problems of current timing layers at HL-LHC, since timing chips usually have high power dissipation.
However, in a high occupancy environment, the event reconstruction with AC-LGADs can be challenging, since each hit cluster extends to several channels.

On the other hand, DC-LGAD technologies (TI-LGADs, DJ-LGADs) limit the cluster size and can tolerate a high occupancy environment, but introduce non-zero dead regions between adjacent channels. This would reduce the fill factor especially in high density pixelizations. 
Furthermore the channel density to achieve a high position resolution would impose a limit on the power consumption per channel of timing chips.
Double-sided LGADs have a higher hit precision and can detect the angle of the particle, however they have a thicker bulk (resulting in reduced time resolution) and require additional readout channels.
In environments with high radiation doses, Buried LGADs, thin LGADs or 3D sensors are more robust against radiation damage and have an advantage. 3D sensors are less affected by radiation damage, however the lower material budged introduced by LGADs make them a more suitable choice for low mass trackers.
Finally, it needs to be noted that any sensor technology requires its integration with the electronics readout. Therefore, power consumption becomes a another key aspect when considering the material budget for future 4D tracking detectors.

\FloatBarrier

\section{Electronics}
%Ariel Schwartzman, Simone Mazza, + additional contributors
%Challenges: power, bandwidth, noise, small area, resolution

While readout prototypes for the timing detectors at the HL-LHC upgrades have demonstrated performance in line with requirements, applying similar techniques in trackers presents several challenges.
High granularity requirement of future trackers will require readout ASICs with smaller pixel sizes compared to present generation, maintaining power consumption levels similar to present designs without timing extraction.
Accommodating the additional required electronics for timing extraction, i.e. Time to Digital Converters (TDCs)~\cite{Angelo} and memories together with the typical pixel circuitry of present trackers, in pixels at pitches on the order of tens of microns, will require the adoption of deeper low power and fast technology nodes beyond 65~nm. 
The entire pixel electronics will need to be designed with low power techniques and with novel timing extraction architectures. In addition, the high luminosity of future hadron colliders will require trackers capable to survive in extreme radiation environments (accumulating a dose of up to 30~GRad and $10^{18}$~neutrons/cm$^2$) 
Because of these aspects, state-of-the-art low power CMOS and Bi-CMOS technology targeted for the mmW communication industry are of particular interest. These includes FDSOI technologies which could potentially open a path to monolithic readouts at very fine pitch. These technologies are also of interest in other HEP applications for their demonstrated performance at deep cryogenic temperatures.
%Advances in detector technology and the direction of HEP experiments and applications require the development of new specialized readout electronics. 
Experimental demands include some combination of high rep rates (order of ns dead time), below 10 ps time of arrival (TOA) resolution, low power (between 0.1 mW and 1 mW per channel), and high dynamic range (for some specific application up to a few thousands, like in the case of PIONEER, see Ref. ~\cite{PIONEER}). 

It is important to stress that the read-out philosophy of standard silicon sensors and LGADs is different. In silicon sensors, the maximum current happens just after the passage of the particle while in LGADs the current increases for the duration of the electron drift time, then there is a plateau, and finally decreases. This peculiar signal shape limits the useful bandwidth of the amplifier. The amplifier bandwidth affects both noise and slope and, ideally, the higher BW the lower the jitter. However, the intrinsic time response of LGAD sensors sets the upper limit to the maximum reachable slope that the analog output can exhibit. As a consequence, the bandwidth should be chosen to be the minimum value that retains the intrinsic sensor speed while keeping the noise low. The bandwidth defines the signal shaping of the front-end and its optimum value for timing is obtained when the amplifier shaping time equals the sensor peaking time.

\subsection{HL-LHC timing chips}

The timing ASICs under development for the ATLAS and CMS timing upgrades, named ALTIROC and ETROC, represent revolutionary steps forward as the first readout chips to bring O(10 ps) timing to collider experiments. However, they are able to use significantly more space and power than high density ASICs designed for trackers with fine pitch and limited material.
In Tab.~\ref{table:timing_chips}, we consider the specifications for these timing chips as well those for RD53A, the ASIC prototype developed for pixel detectors at the HL-LHC. Compared to RD53A, the timing chips use several hundred times more power and area per channel. The requirements for radiation tolerance, on the other hand, are more similar (and are merely the minimum requirements, not necessarily the upper limit). Indeed, the primary challenges to transform the timing chips into chips for 4D tracking will be to minimize both the power consumption and the channel size.

\begin{table}[htp]
  \centering
  \caption{Requirements for state-of-the-art readout chips designed for timing (ALTIROC~\cite{CERN-LHCC-2020-007} and ETROC~\cite{ CMS:2667167}) and for pixel detectors (RD53A / CMS Phase II tracker \cite{Garcia-Sciveres:2287593,ref:mucol:CMSupgrade}.)}
  \begin{tabular}{ c | c | c | c | c | c}
    ASIC        & Technology & Pitch & Total size & Power consumption & TID tolerance  \\
%   & &  [\si{\milli\m}] &  [\si{\milli\m^2}] &[mW/chan] & [MRad] \\
 \hline
  ALTIROC         & 130 nm   & \SI{1.3}{\milli\m}  & $19.5 \times 19.5~\si{\milli\m^2}$& 5 \si{\milli \W}/chan &  2 MGy \\ 
  ETROC         & 65 nm   & \SI{1.3}{\milli\m}  & $20.8 \times 20.8 ~\si{\milli\m^2}$ & 3 \si{\milli \W}/chan  & 1 MGy \\ 
  RD53A/HL-LHC pixels  & 65 nm & \SI{50}{\micro\m} & $20 \times 11.6 ~\si{\milli\m^2}$ & $<10$ \si{\micro \W}/chan & 5--15 MGy \\ \hline

  %\hline
  \end{tabular}
  \label{table:timing_chips}
\end{table}

\subsection{Prototype timing chips}
Currently there are many ongoing projects with the aim of improving the available timing chips. Efforts are geared towards a specific application, however the goals are common: bandwidth optimization, low noise, low area, time resolution and power dissipation. Following is a non complete list of current efforts:
\begin{itemize}
    \item CERN’s EP R\&D WP5: CMOS Technologies~\cite{CERN-EP-RDET-2021-001} survey has promoted the selection of 28~nm CMOS node as the next step in microelectronics scaling for HEP designs. The 28~nm technology is at least twice as fast and allows circuit densities around 4-5 times higher than the previously employed 65~nm node, making it a good candidate for design of high granularity 4D trackers. One of the critical circuit blocks necessary to enable 4D operation in trackers are low-power and compact TDCs capable of high time-measurement precision. SLAC has started the design of TDCs in 28nm technology node with target time resolutions of 10-50ps. The plan is to submit the first prototype for fabrication at the end of this year.

    \item FAST family of ASICs~\cite{FAST2}: based on 110 nm CMOS commercial technology, optimized for the readout of LGAD sensors. The channel architecture consists of a Trans-Impedance Amplifier, a second amplification stage based on a common source amplifier, and a two-stage leading edge discriminator. Each channel can measure the Time of Arrivals (ToA) and Time of Threshold (ToT) of a pulse signal with a least significant bit of 25~ps. 
    \item Silicon-Germanium: a possible path to achieve O(10~ps) time resolution is an integrated chip using Silicon Germanium (SiGe) technology. Prototype SiGe front end readout chip optimized for low power and timing resolution are being produced by Anadyne, Inc. in collaboration with the University of California Santa Cruz. The chip is expected to have 0.5~mW per channel (front end and discriminator) while retaining 10~ps of timing resolution for 5~fC of injected charge. The production is made with TowerJazz 130~nm process.
    \item Full digitization chip: University of California Santa Cruz is currently working with Nalu Scientific to design and fabricate a high channel density and scalable radiation-hard waveform digitization ASIC with embedded interface to AC-LGAD sensor arrays. The chip was manufactured with TSMC’s 65nm technology. It is a full digitization chip with full waveform digitization, which is expected to be more robust against a variety of adverse factors which can affect timing and spatial precision of AC-LGADs.
\end{itemize}
Additional details on the above cited efforts can be found in the "electronics for fast timing" snowmass paper~\cite{TimingElectronics}.

\section{Layout}
%Ariel Schwartzman, Ryan Heller, + additional contributors

A major next step towards 4D tracking at future hadron colliders is the study of how to best combine timing with spatial information. The fine spatial tracking resolution demand towards small pixel with low material budget and low power may make it impractical to instrument finest timing capabilities on all layers. On the other hand, 4D devices with still fine spatial granularity and integrated some coarse timing capabilities can potentially allow a versatile mixture of layers with different balance of spatial and timing resolution to serve an optimal overall 4D tracking for the wide range of applications. 
The addition of timing information to every pixel hit might not be the approach that leads to the best performance. Alternative approaches such as alternating spatial with timing layers, or 4D with 3D layers could help improve the overall physics performance. 

Another aspect of detector layout is related to the physics drivers motivating its development. For example, improved and fast charged track reconstruction, heavy flavor ($b/c$) tagging, and particle-flow reconstruction under very high pileup density will require 4D information in the inner layers, whereas LLP and time-of-flight particle ID capabilities, including the possibility of strange-tagging~\cite{Abe_2000, nakai2020strange, Erdmann_2020, sLoI}, will benefit from 4D information in the outer layers. LLP applications would demand continuous timing coverage and could benefit from modest timing resolution in more layers without stretching timing dynamic range. Future $e^+e^-$ collider vertex detector backgrounds are predominantly back-scattered bremsstrahlung particles from downstream magnets and collimators with $\sim$ns range delays. 4D tracking devices with fine spatial resolution and modest timing resolution in other layers could significantly enhance the overall performance. 

%\textcolor{Mention EIC layout? AC-coupling -> lower density?}

Other key considerations are tracking material and pseudorapidity coverage. The additional material required to go from 3D to 4D tracking will have an impact on the track-time association efficiency and mis-association rate. Whereas a lower track-time efficiency will simply reduce the potential gains from timing information, the wrong assignment of times to tracks is particularly problematic as in this case the 4D reconstruction will perform worse than 3D. The impact of showering of particles within the tracking material might be partially mitigated with the use of advanced algorithms based on graph neural networks or other deep learning techniques but this will require a long term study. 
In the case of future lepton colliders, material in the tracking detector has to be minimized to not degrade $p_t$ and impact parameter resolution, posing additional constrains on the incorporation timing information~\cite{1306.6329} 

\section{Key areas for future R\&D}

Having presented the case for timing at future collider experiments and outlining the technologies considered to meet these needs, we summarize in this section the key research directions needed to realize 4D trackers.

Although much progress has been accomplished recently in 4D sensor design, critical areas of development remain. The first step to achieve 4D capability from timing sensors, introducing fine segmentation, has already been demonstrated with the introduction of AC-LGADs. These sensors are well-suited in particular for applications with low to moderate occupancy and enable very fine spatial resolution with sparse channel density. However, AC-LGADs face challenges in extreme occupancy environments, where isolation between channels becomes important. Future work should also be devoted to other innovative segmentation techniques that isolate adjacent channels, including trench isolation (TI-LGADs), which have already begun to be demonstrated, and deeper gain layer structures (DJ-LGADs) that allow fine segmentation without AC-contacts.

Beyond fine segmentation, developing sensors with adequate radiation tolerance continues to be the most significant challenge. Several different approaches should be pursued to address this issue, including strategies for deeper gain layers or more sophisticated gain structures, as in buried LGAD designs. Thinner sensor designs should also be developed, which enable full depletion even at extreme fluences as high as $10^{17}$ neq/cm$^2$. Substantial R\&D effort must be devoted to improving sensor radiation hardness in order to build a tracker adequate for a future hadron collider.
%\textcolor{red}{Maybe could reference facilities at this stage (reactors, accelerators for proton irradiation, accelerators for beam tests, SIMS / imaging techniques for material properties}

Available timing chips developed for the ATLAS and CMS timing layers allow to reach a time resolution of O(10~ps). However, compared to standard pixel readout (e.g. RD53A), they have several hundred times more power and area per channel.
Therefore to achieve true 4D tracking, O(10~ps) and O(10~$\mu$m), the primary challenge for readout chips will be to minimize both the power consumption and the channel size.
On top of this requirements the TID sustained in future hadron colliders will be much higher than current specifications.
Community efforts aimed to timing chip development for specific applications need to have the same common goals: bandwidth optimization, radiation hardness, low noise, low area and low power dissipation while maintaining great time resolution.

Achieving these R\&D goals will critically leverage dedicated DOE tests facilities at accelerators for beam tests, and proton irradiation studies. These facilities are a necessity and must be well supported. Examples include: 
\begin{itemize}
            \item Protons, pion, and electron test beam facilities to characterize sensor and chip performance and resolution
            \item Irradiation facilities providing 1~MeV neutrons, protons at various energy, and gamma rays for radiation damage characterization. Intensity has to be high enough to reach $10^{18}$~Neq.
            \item Facilities providing a high ionization environment (low energy ion beams and high energy lasers) to study LGAD secondary properties such as Single event burnout (SEB), and gain suppression. 
            \item Secondary Ion Mass Spectrometry (SIMS) to study silicon sensor structures
\end{itemize}
        
In addition, new DOE facilities such as high-repetition, high-intensity, free electron lasers may provide new tools for imaging and characterization of devices at extreme conditions.

\section{Summary}
Advanced 4-dimensional trackers with ultra fast timing (10-30 ps) and extreme spatial resolution (O(few $\mu m$)) represent a new avenue in the development of silicon trackers, enabling a variety of physics studies which would remain out of experimental reach with the existing technologies. 

In this paper, we described the impact of integrating 4D tracking capabilities both in potential upgrades of the LHC experiments and in several detectors at future colliders. The physics drivers vary, along with distinct features of hadron or lepton machines, resulting in stringent constraints on time and space resolution, material budget and radiation hardness. 
We summarized the key components that determine the final time resolution in a detector and the ideal features of such a device. The time walk and Landau resolution components, intrinsic to the sensor, can be reduced respectively with short drift time and limited thickness in the path of a MIP. The time errors arising from the jitter and the TDC, which are instead related to the readout chip’s electronics, benefit respectively from high signal to noise ratio and small TDC bin size. 
We have then reviewed and compared state-of-the-art silicon technologies.
A large effort is ongoing on leveraging the good timing performance of Low Gain Avalanche Detectors and adapting them to achieve fine segmentation. While work is still needed on improving radiation tolerance, promising approaches emerge, which could achieve space resolutions below 10 $\mu m$. 
Alternatively, tracking sensors can be modified to facilitate precision timing. This could be achieved by exploiting 3D geometries with short drift times or through closer integration of the sensor with the electronics, for example via induced current or monolithic detectors.

The high granularity requirements at future trackers will also pose unprecedented constraints on the readout electronics for hybrid detectors: the ASICs will have to be integrated on smaller pixel sizes compared to the present generation while maintaining power consumption levels comparable to ASICs designs without timing extraction. At the same time, refinements in monolithic detectors can make them suitable for fast timing applications.

We identified key areas for future Research \& Development which will lead us towards the deployment of 4D (or even 5D) tracking and potentially also spin-off activities. These range from sensor and electronics design to the realization of dedicated test beams and irradiation campaigns.

Bringing in technological innovation and fully exploiting the potential of future detectors through the usage of ultra fast timing is a unique and exciting opportunity for the particle physics community. In order to reach this goal, it is of paramount importance in the coming years to undertake thoroughly the R\&D studies mentioned above and to investigate how to best combine timing with spatial information, both in terms of modern detector layouts and via innovative reconstruction algorithms.

\section*{Acknowledgements}

The authors of this paper would like to thank the SLAC TID group, in particular Larry~Ruckman (SLAC) for the fruitful discussions on front-end and readout electronics. The authors are also grateful to Pierfrancesco~Butti (SLAC), Markus~Elsing (CERN), Fabian~Klimpel (CERN) and Andreas~Salzburger (CERN) for the brainstormings about the usage of precision time in track reconstruction, to Valerio~Dao (CERN) for his helpful insights on the MALTA developments and to Su~Dong (SLAC) for the countless constructive dialogues from detector features to physics applications.

\clearpage

\bibliographystyle{spphys} % APS-like style for physics
\bibliography{bibliography}

\end{document}